\documentclass[useAMS,usenatbib]{mn2e}

\usepackage[psamsfonts]{amssymb}
\usepackage[dvips]{graphicx}
\usepackage{amsmath,alltt}
\usepackage{multirow}
\usepackage{rotating}
\usepackage{lscape}

\title[The Origins of BSs and Binarity in GCs]{The Origins of Blue Stragglers and Binarity in Globular Clusters}  

\author[Nathan Leigh, Christian Knigge, Alison Sills, Hagai Perets,
Ata Sarajedini, and Evert Glebbeek]{Nathan
  Leigh$^{1,2}$, Christian Knigge$^{3}$, Alison Sills$^{1}$, Hagai B. Perets$^{4}$,
\newauthor
Ata Sarejedini$^{5}$, Evert Glebbeek$^{1,6}$\thanks{E-mail:
    leighn@mcmaster.ca (NL); asills@mcmaster.ca (AS1);
    hperets@cfa.harvard.edu (HP); christian@astro.soton.ac.uk (CK);
    ata@astro.ufl.edu (AS2); e.glebbeek@astro.ru.nl (EG)} \\
$^{1}$Department of Physics and Astronomy, McMaster University,
1280 Main St. W., Hamilton, ON, L8S 4M1, Canada \\
$^{2}$European Space Agency, Space Science Department, Keplerlaan 1,
2200 AG Noordwijk, The Netherlands \\
$^{3}$School of Physics and Astronomy, University of Southampton,
Highfield, Southampton, SO17 1BJ, United Kingdom \\
$^{4}$Harvard-Smithsonian Center for Astrophysics, 60 Garden
St., Cambridge, MA 02338, USA \\
$^{5}$Department of Astronomy, University of Florida, Gainesville, FL
32611, USA \\
$^{6}$Department of Astrophysics/IMAPP, Radboud University Nijmegen,
P.O. Box 9010, 6500 GL Nijmegen, The Netherlands}
\begin{document}

\pagerange{\pageref{firstpage}--\pageref{lastpage}} \pubyear{2012}

\maketitle

\label{firstpage}

\begin{abstract}

Two basic formation channels have been proposed for blue straggler
stars in globular clusters: binary star evolution and stellar
collisions. We recently showed that the number of blue stragglers 
found in the core of a globular cluster is strongly correlated with
the total stellar mass of the core, but not with the collision rate in
the core. This result strongly favoured binary evolution as the dominant
channel for blue straggler formation. Here, we use newly available
empirical binary fractions for globular clusters to carry out a more
direct test of the binary evolution hypothesis, but also of
collisional channels that involve binary stars. More specifically, using
the correlation between blue straggler numbers and core mass as a
benchmark, we test for correlations with the number of binary stars,
as well as with the rates of single-single, single-binary, and
binary-binary encounters. We also consider joint models, in which blue
straggler numbers are allowed to depend on star/binary numbers and
collision rates simultaneously. 

Surprisingly, we find that the simple 
correlation with core mass remains by far the strongest predictor of
blue straggler population size, even in our joint models.  
This is despite the fact that the binary fractions
themselves strongly anti-correlate with core mass, just as expected in
the binary evolution model.

At first sight, these results do not fit neatly with either binary
evolution or collisional models in their simplest forms.  
Arguably the simplest and most intriguing possibility to explain 
this unexpected result is that
observational errors on the core binary fractions are larger
than the true intrinsic dispersion associated with their dependence on
core mass. In the context of the binary evolution model, this would
explain why the combination of binary fraction and core mass is a
poorer predictor of blue straggler numbers than core mass alone. It would also
imply that core mass is a remarkably clean predictor of core binary
fractions. This would be of considerable importance for the dynamical 
evolution of globular clusters, and provides an important benchmark 
for models attempting to understand their present-day properties.

\end{abstract}


\section{Introduction} \label{intro6}

Blue stragglers (BSs) in globular clusters (GCs) are stars that appear
brighter and  
bluer than the main-sequence turn-off (MSTO) in the cluster 
colour-magnitude diagram (CMD) \citep{sandage53}. They are 
thought to be created 
when fresh hydrogen is mixed into the core of a normal low-mass 
main-sequence (MS) star \citep[e.g.][]{sills01,sills02}. Two main 
formation channels have been 
proposed for BSs in GCs: binary star 
evolution and dynamical interactions. BSs can form via the 
former pathway either through binary mass-transfer due to Roche-lobe
overflow from an evolved primary onto a normal  
MS companion \citep{mccrea64,geller11}, or through the coalescence 
of two normal MS stars in a binary system. The latter mechanism can 
occur due to angular momentum loss induced 
by a magnetized stellar wind \citep[e.g.][]{iben84,
 andronov06}, or even Kozai cycles induced by an outer triple 
companion \citep{perets09}. The dynamical 
pathway involves collisions between two or more MS stars 
\citep[e.g.][]{sills99}. These 
are typically mediated by dynamical interactions involving 
binary stars, since the cross-section for collision is much 
larger for a binary than it is for a single star 
\citep[e.g.][]{leonard89, leonard92}. 

Several statistical studies have been conducted in search of 
a dominant BS formation channel. However, the 
cluster parameter that has thus far yielded the
strongest correlation with 
BS population size is the cluster \citep{piotto04,leigh07}
or core \citep{knigge09,leigh11a} mass. Many authors have tried to
explain this by simultaneously invoking multiple formation 
mechanisms. For example, \citet{davies04} suggested that the 
observed dependence of BS numbers on total cluster mass can be 
explained if BSs in low-mass clusters are
primarily descended from binaries, whereas BSs in high-mass
clusters are primarily descended from collisions. A similar
scenario has been argued for in an attempt to explain the
bimodal BS radial distribution observed in many globular 
clusters 
\citep[e.g.][]{ferraro93,ferraro04,mapelli06,lanzoni07,beccari11,sanna12}.
In this picture, BSs in the dense core were formed in collisions, 
whereas BSs in the low-density cluster outskirts were
formed by mass-transfer within primordial binaries.

At the time these studies were conducted, there were hardly any
observational constraints on the properties of the binary populations
in GCs. In particular, empirical binary fractions were available for
only a small subset of low-density GCs \citep{sollima08}. This issue was only recently 
resolved by \citet{milone11}. Using data from the HST-based ACS Survey  
of Globular Clusters, these authors derived photometric 
binary fractions for the MS populations in 59 Milky Way (MW) 
GCs. This sample offers a long-awaited opportunity to test more
directly whether the sizes of BS and binary populations in GCs are
correlated, as one might expect for both the binary evolution channel
and for collisional formation channels that involve binaries (i.e. 1+2
and 2+2 encounters). This was previously addressed by \citet{sollima08} 
and \citet{milone11}.  These results provided evidence that the 
blue straggler fraction is indeed related to the binary fraction, and 
we build on those previous works in this paper.

The plan of this paper is as follows. First, in Section~\ref{data}, we
present the observational data. We then explain and
carry out our analysis of these data in Section~\ref{analysis}. There,
we derive the expected scaling laws for the simplest versions of the
various formation  
channels and compare these theoretical predictions to the
observations. Finally, the implications of our results for the
formation and evolution of BSs in GCs are discussed in
Section~\ref{discussion}.

\section{Observational Data} \label{data}

BS numbers are taken from Table 1 of \citet{leigh11a}, which was 
compiled using data taken from the ACS Survey for Globular 
Clusters \citep{sarajedini07}\footnote[1]{The data can be found at
http://www.astro.ufl.edu/$\sim$ata/public\_hstgc/, and was last
accessed on 02/02/11.} The sample used in this paper omits 
five clusters from the catalogue of \citet{leigh11a}, since we 
do not have observed binary fractions in these cases. We use 
only those BS number counts within the core and within four core 
radii from the cluster centre (columns 4 and 7, respectively, in 
Table 1 of \citet{leigh11a}) in this paper. 

Binary fractions within the core (r $<$ r$_c$; f$_{bin,core}$) 
and the half-mass radius (r $<$ r$_h$; f$_{bin,half}$) are 
taken from Table 1 of \citet{milone11}.
\footnote{These binary fractions have been corrected for 
a variety of observational biases, including completeness,
contamination from field stars, and differential reddening. Detailed
explanations of these procedures have been provided in \citet{sarajedini07},
\citet{anderson08} and \citet{milone11}.}
The latter values are not 
provided directly in \citet{milone11}. Instead, binary 
fractions within the annulus separating the core and the half-mass 
radius (r$_c$ $<$ r $<$ r$_h$; f$_{bin,r_c < r < r_h}$) are given. 
Therefore, we calculate mass-weighted binary fractions within the 
half-mass radius using the relation:
\begin{equation}
\label{eqn:f_bin_half}
f_{bin,half} = \frac{f_{bin,core}M_{core} + f_{bin,r_c < r < r_h}(M_{half} - M_{core})}{M_{half}},
\end{equation}
where M$_{core}$ and M$_{half}$ are the mass of the cluster core and 
the mass contained within the half-mass radius, respectively. 
In order to obtain accurate estimates for the total stellar mass
contained within the core, we generated single-mass King models calculated 
using the method of \citet{sigurdsson95} to obtain luminosity density 
profiles for every cluster in our sample. The profiles were found using
the concentration parameters of \citet{mclaughlin05} and the central
luminosity densities of \citet{harris96}. We then integrated the derived 
luminosity density profiles numerically in order to estimate the total
stellar light contained within the core, which we multiplied by a 
mass-to-light ratio of 2 in order to obtain estimates for the total stellar 
mass contained within one core radius from the cluster centre. The mass 
enclosed within the half-mass radius was then estimated by calculating
the total 
cluster mass from its absolute integrated visual magnitude (once again 
assuming a mass-to-light ratio of 2), and then dividing by two. Throughout 
this paper, we 
have adopted the binary fractions provided in column 6 of Table 1 in 
\citet{milone11}, which 
provides the \textit{total} fraction of objects that are binaries 
within the indicated annulus. However, in practice, this estimate of
the total number of binaries is extrapolated from the observed
fraction of binaries with mass ratio $q > 0.5$ by assuming a flat
distribution in $q$.

\begin{figure}
\begin{center}
\includegraphics[width=\columnwidth]{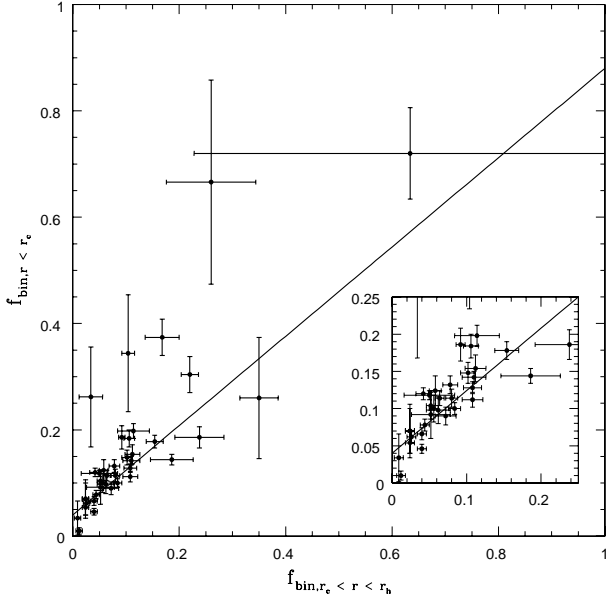}
\end{center}
\caption[The binary fraction in the core plotted versus the binary fraction in
the annulus separating the core and the half-mass radius]{The binary
 fraction in the core plotted versus the binary fraction in
the annulus separating the core and the half-mass radius. Error bars
are taken directly from Table 1 of \citet{milone11}. The solid line
shows the weighted least-squares fit to the data provided in
Equation~\ref{eqn:fbh_vs_fbc}.
\label{fig:fbh_vs_fbc}}
\end{figure}

There are several clusters for which the binary fractions in
\citet{milone11} are provided for f$_{bin,r_c < r < r_h}$ but not
f$_{bin,core}$. To approximate f$_{bin,core}$ in those clusters
for which these values are missing, we perform a weighted least-squares
fit to quantify
the dependence of f$_{bin,core}$ on f$_{bin,r_c < r < r_h}$ using every
cluster in the sample of \citet{milone11} for which both of these
quantities were given. We then supplement f$_{bin,core}$
in every cluster for which only f$_{bin,r_c < r < r_h}$ was provided. This
is only necessary in a handful of clusters, but the resulting increase
in our sample size is nevertheless worth while. We obtain a relation
of the form:
\begin{equation}
\label{eqn:fbh_vs_fbc}
f_{bin,core} = (0.84 \pm 0.15)f_{bin,r_c < r <r_h} + (0.04 \pm 0.01).
\end{equation}
This is shown in Figure~\ref{fig:fbh_vs_fbc}. The fit is good for
binary fractions less than $\sim 0.2$, but the agreement is poor
for larger binary fractions. In order to test the effects
had on our least-squares fit by clusters with large binary fractions, we
re-perform it considering only those clusters for which both
f$_{bin,core} < 0.2$ and f$_{bin,r_c < r < r_h} < 0.2$. In this case,
both the slope and y-intercept of the fit agree with those reported
in Equation~\ref{eqn:fbh_vs_fbc} to within one standard deviation.
We conclude that the fit and the corresponding uncertainties provided in
Equation~\ref{eqn:fbh_vs_fbc} are reasonable.

Our analysis in Section~\ref{analysis} also requires the absolute visual
magnitude (M$_V$), core radius (r$_c$), central luminosity density
($\rho_0$) and central velocity dispersions ($\sigma_0$) of each
cluster. All of these quantities are taken from \citet{harris96},
except for the velocity dispersion of 10 clusters for which
these values are not provided. In these cases, we use the calculated
velocity dispersions provided by \citet{webbink85}.

\section{Analysis and Results} \label{analysis}

In this section, we derive theoretical scaling laws for the simplest
versions of the various proposed blue straggler formation channels and
compare these theoretical predictions to the observational data. In
practice, this means we will search for correlations between the
number of blue stragglers in the core of each GC and the cluster
parameter that should set this number in each scenario. 

We assess the significance of our correlations primarily via the
Spearman rank test, which provides both a correlation coefficient
(r$_s$) and the significance level at which 
the null hypothesis of zero correlation is disproved (p$_s$). A small
$p_s$-value is indicative of a significant correlation. In addition, 
we perform weighted least-squares fits to quantify the dependence of 
BS numbers on each parameter. Uncertainties for all number counts are
obtained assuming Poisson statistics, but we also allow for intrinsic
dispersion in our fits, at whatever level is required to achieve a
reduced ${\chi}^{2} \simeq 1$.

Table~\ref{table:least-squares} shows the results of our comparisons
between the observed BS numbers and the parameter we have tested. 
Each entry in Table~\ref{table:least-squares} gives the slope for the
line of best-fit, Spearman correlation coefficient, and probability at
which the null hypothesis of zero correlation is disproved. These
are provided in the form (slope; r$_s$, p$_s$).

\subsection{Core Mass} \label{core}

We begin by revisiting the correlation between core BS numbers and core 
mass, with the latter estimated from our King model fits. Core mass
was found to be 
the best predictor of BS numbers in the core in \citet{knigge09} and
\citet{leigh11b}. Our working assumption in those studies was that
core mass was able to predict BS numbers because it is a partial 
proxy for the (unknown) number of binaries in the core (since
$N_{bin,core} = f_{bin,core} M_{core}$; see
Section~\ref{binaries}). Now that binary fractions are available, we 
can test this assumption directly. However, the original correlation
with just core mass remains a useful benchmark in this context:
if the empirical binary fractions have added useful information,
including them should allow us to discover even stronger correlations.

\begin{figure}
\begin{center}
\includegraphics[width=\columnwidth]{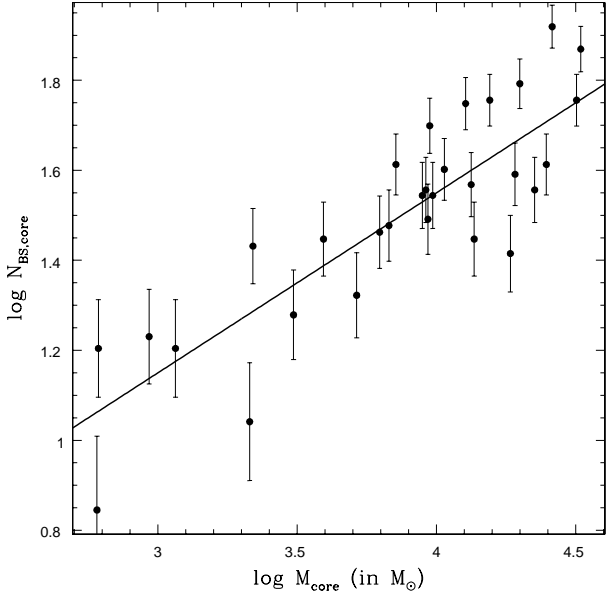}
\end{center}
\caption[The logarithm of the number of BSs as
a function of the core mass]{The
 logarithm of the number of
BSs as a function of the core mass (in units of solar
masses). The solid line shows the best-fit to the data.
\label{fig:Mcore_vs_Npop}}
\end{figure}

We find a dependence $N_{BS,core} \propto M_{core}^{0.40 \pm 0.05}$, 
and a Spearman correlation coefficient 0.83. This correlation is shown in 
Figure~\ref{fig:Mcore_vs_Npop}, along with the corresponding line of
best-fit to the data. The power-law index on
$M_{core}$ is inconsistent with zero at the $8\sigma$ confidence level, 
but also inconsistent with unity at the $12\sigma$ confidence level. 
Therefore, the strong, sub-linear correlation between $N_{BS,core}$
and $M_{core}$ first reported in \citet{knigge09} is confirmed in the
present analysis.

\subsection{Binary Population Size} \label{binaries}

If most of the BSs in our sample are descended from binary evolution,
we predict a dependence of the form:
\begin{equation}
\label{eqn:N_pop_bin}
N_{BS,core} \propto N_{bin,r} \sim \frac{f_{bin,r}M_{encl,r}}{\bar{m}},
\end{equation}
where $N_{BS}$ is the number of BSs within a given radius $r$ from
the cluster centre, $N_{bin,r}$ is the number of binaries contained within
$r$, $f_{bin,r}$ is the fraction of objects within $r$ that are binaries,
$M_{encl,r}$ is the total stellar mass enclosed within $r$, and $\bar{m}$ 
is the average stellar mass (for which we assume the same value in all
clusters). 

As pointed out in \citet{knigge09}, Equation~\ref{eqn:N_pop_bin}
allows a simple explanation for the observed sub-linear correlation
between $N_{BS,core}$ and $M_{core}$. If we identify $M_{core}$ with
$M_{encl,r}$ (i.e. we take the relevant radius to be $r = r_{core}$),
then $N_{BS,core} \propto M_{core}^{0.4}$ follows from
Equation~\ref{eqn:N_pop_bin} if the core binary fraction is itself a
function of core mass, i.e. $f_{bin,core} \propto M_{core}^{-0.6}$. We
also showed in \citet{knigge09} that there was indeed some evidence
for an anti-correlation between core mass and core binary frequency,
based on the data set of \citet{sollima08}; this is a prelimimary 
version of the ACS-based data set we use here.

\begin{figure}
\begin{center}
\includegraphics[width=\columnwidth]{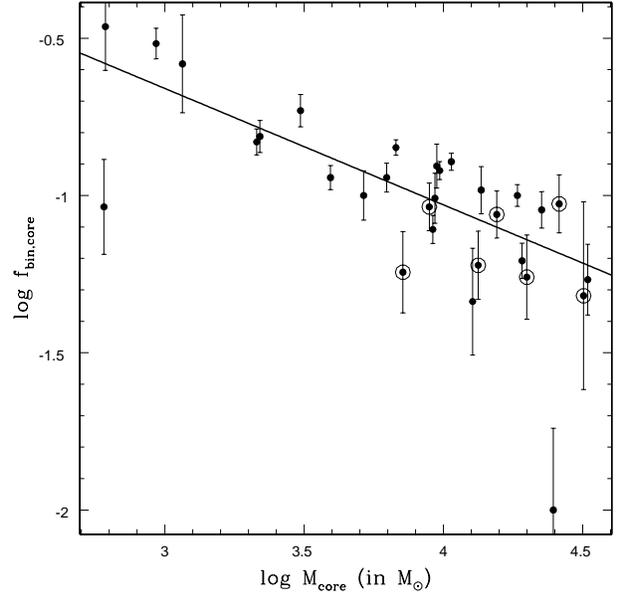}
\end{center}
\caption[The logarithm of the core binary fraction as
a function of the core mass]{The logarithm of the core binary fraction
as a function of the core mass (in units of solar masses). The solid
line shows the best-fit to the data. The seven data points with open
circles around them correspond to the core binary fractions calculated
using Equation~\ref{eqn:fbh_vs_fbc}.
\label{fig:Mcore_fbin}}
\end{figure}

The final ACS binary fraction data set available now \citep{milone11} 
allows us to check whether the evidence for this anti-correlation
holds up. Figure~\ref{fig:Mcore_fbin} shows that it
does. More specifically, we find $f_{bin,core} \propto M_{core}^{-0.37
  \pm 0.06}$, while the Spearman correlation coefficient is r$_s =
-0.72$. The slope for this anti-correlation is not quite as steep as
in our naive prediction, but the existence of the anti-correlation
itself is obviously promising. 

\begin{figure}
\begin{center}
\includegraphics[width=\columnwidth]{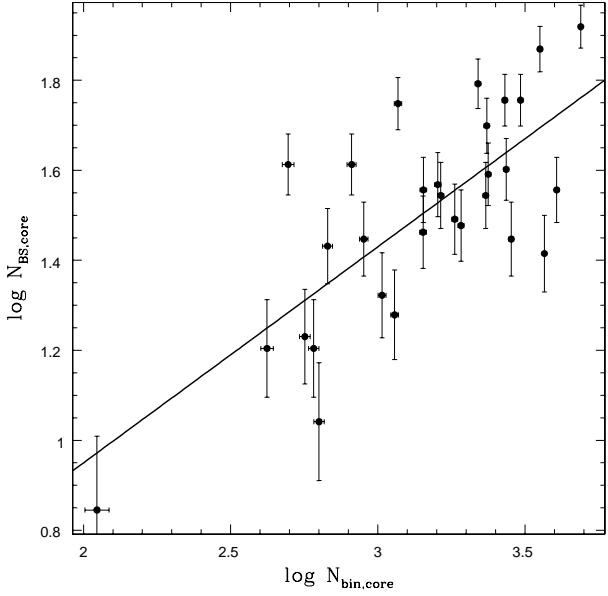}
\end{center}
\caption[The logarithm of the number of BSs as
a function of the logarithm of the number of binary stars in the
core]{The logarithm of the number of
BSs (within the core only) as a function of the
number of binary stars in the core.
The solid line shows the best-fit to the data (the slope for which
is shown in Table~\ref{table:least-squares}).
\label{fig:Nbc_vs_Npop}}
\end{figure}

\begin{figure}
\begin{center}
\includegraphics[width=\columnwidth]{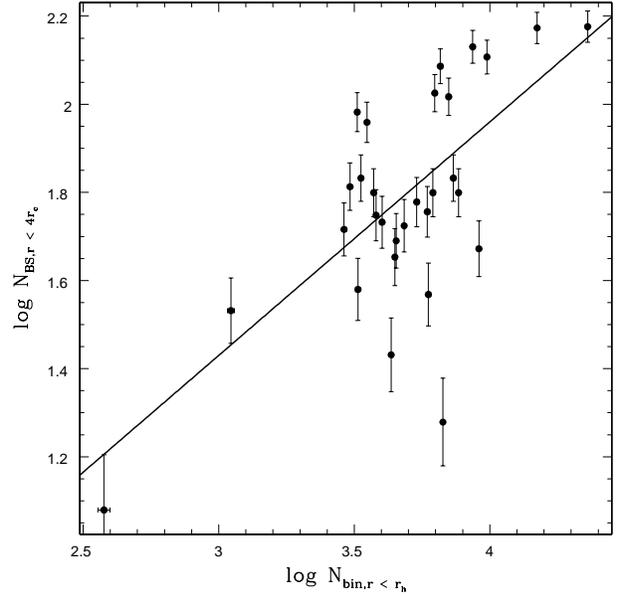}
\end{center}
\caption[The logarithm of the number of BSs as
a function of the number of binary stars within the
half-mass radius]{The logarithm of the number of
BSs (within four core radii from the cluster centre)
as a function of the number of
binary stars within the half-mass radius.
The solid line shows the best-fit to the data.
\label{fig:Nbh_vs_Npop}}
\end{figure}

However, this promise is not actually borne out. When we compare 
$N_{BS,core}$ directly to $N_{bin,core} = f_{bin,core}M_{core}$
(Figure~\ref{fig:Nbc_vs_Npop}; Table~\ref{table:least-squares}), we
find that the {\em strength} of the 
correlation actually {\em decreases} compared to the correlation with
just $M_{core}$ (the Spearman rank coefficient drops from 0.83 to
0.61), while the best-fit {\em slope} increases
only marginally (from $0.40 \pm 0.05$ to $0.48 \pm 0.09$). Increasing 
the size of the region considered does not improve things: when we
compare the number of BSs within four core radii from the cluster 
centre against the number of binaries within the half-mass radius, we
still find only a weak correlation with a clearly sub-linear slope 
(Figure~\ref{fig:Nbh_vs_Npop}; Table~\ref{table:least-squares}).
This is a surprising result at first sight, and we will discuss its
implications further in Section~\ref{discussion}. 

Before moving on, it is worth stressing that the strong
anti-correlation between binary fraction and core masse shown in
Figure~\ref{fig:Mcore_fbin} is interesting in its own
right. Empirically, its existence  
is in line with a similarly strong correlation between core binary
fraction and cluster absolute magnitude that was already presented in 
\cite{milone11}). This is because absolute magnitude is a proxy for
total cluster mass, which in turn correlates strongly with core mass
among GCs. Theoretically, however, it is far from clear why abundance
of binary stars should depend so strongly on either the core or the
total stellar mass of their host clusters. The fact that it does must
be important for our understanding of cluster dynamics. Presumably,
dynamical evolution is responsible for establishing this correlation
and is, in turn, affected by it. 

\subsection{Collision Rates} \label{collisions}

If most of the BSs in our sample were formed from direct collisions
between single stars, then we predict a dependence of the form:
\begin{equation}
\label{eqn:N_pop_coll}
N_{BS} \propto N_{1+1} \sim \int_{\tau_0}^{\tau_{cl}}\Gamma_{1+1}dt,
\end{equation}
where $\Gamma_{1+1} = 1/\tau_{1+1}$ is the rate of single-single collisions
producing BSs (we use the form given in \citet{leigh11b}), and we are integrating with respect to time.  We use the age of
the cluster ($\tau_{cl}$) as the upper limit of integration, and the
average BS age ($\tau_{BS}$) to calculate the lower limit of integration
according to $\tau_0 = \tau_{cl} - \tau_{BS}$.  Assuming $\Gamma_{1+1}$
remains constant in time \citep{leigh11c}, Equation~\ref{eqn:N_pop_coll} simplifies to:
\begin{equation}
\label{eqn:N_pop_coll2}
N_{BS} \propto N_{1+1} \sim \Gamma_{1+1}\tau_{BS}.
\end{equation}

If most of the BSs in our sample were formed from dynamical interactions
involving binaries, then we predict either a relation of the form:
\begin{equation}
\label{eqn:N_pop_1+2}
N_{BS} \propto N_{1+2} \sim \Gamma_{1+2}\tau_{BS},
\end{equation}
where $\Gamma_{1+2} = 1/\tau_{1+2}$ is the rate of single-binary
encounters, or:
\begin{equation}
\label{eqn:N_pop_2+2}
N_{BS} \propto N_{2+2} \sim \Gamma_{2+2}\tau_{BS},
\end{equation}
where $\Gamma_{2+2} = 1/\tau_{2+2}$ is the rate of binary-binary encounters.

\begin{figure}
\begin{center}
\includegraphics[width=\columnwidth]{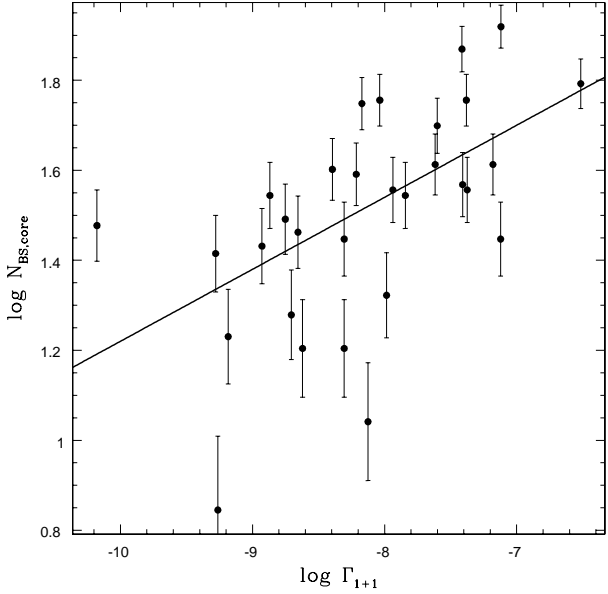}
\end{center}
\caption[The logarithm of the number of BSs as
a function of the single-single collision rate in the core]{The
  logarithm of the number of
BSs as a function of the single-single collision
rate in the core (in units of number of collisions per year).  The
solid lines shows the best-fit to the data.
\label{fig:Gamma11_vs_Npop}}
\end{figure}

\begin{figure}
\begin{center}
\includegraphics[width=\columnwidth]{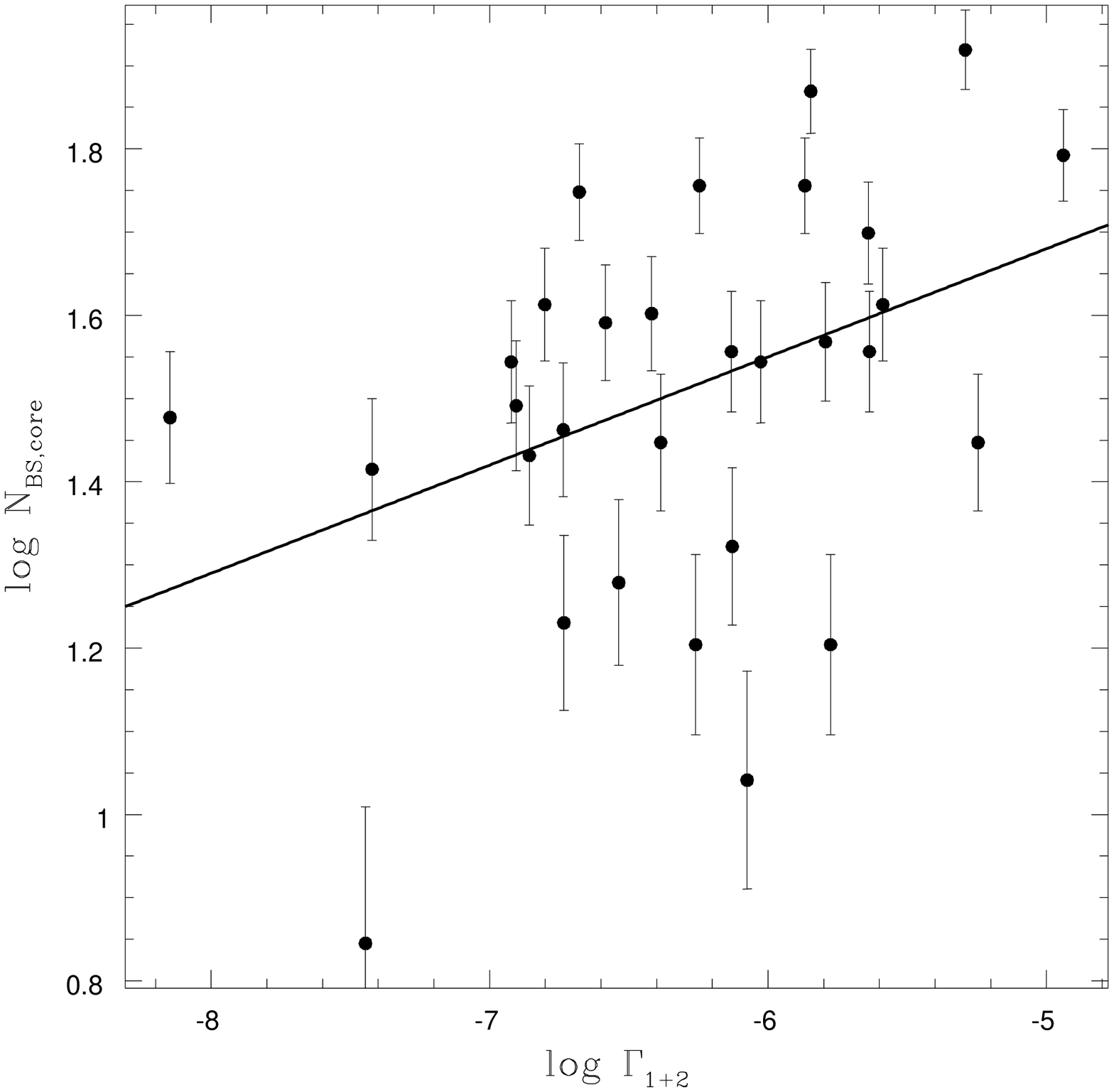}
\end{center}
\caption[The logarithm of the number of BSs as
a function of the single-binary collision rate in the core]{The
  logarithm of the number of
BSs as a function of the single-binary collision
rate in the core (in units of number of collisions per year).  The
solid line shows the best-fit to the data.
\label{fig:Gamma12_vs_Npop}}
\end{figure}

\begin{figure}
\begin{center}
\includegraphics[width=\columnwidth]{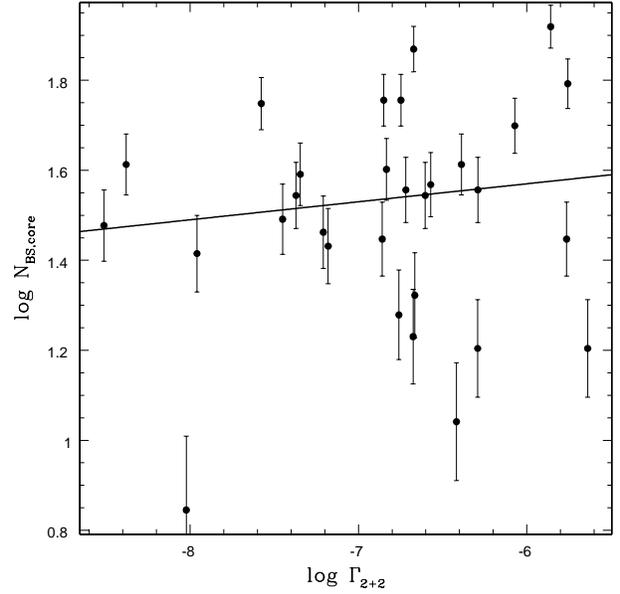}
\end{center}
\caption[The logarithm of the number of BSs as
a function of the binary-binary collision rate in the core]{The
  logarithm of the number of
BSs as a function of the binary-binary collision
rate in the core (in units of number of collisions per year).  The
solid line shows the best-fit to the data.
\label{fig:Gamma22_vs_Npop}}
\end{figure}

We find a correlation between the number of BSs in the core 
and the 1+1 collision rate.  The slope
for this relation is inconsistent with zero at nearly the $4\sigma$
confidence level, and it yields a Spearman correlation coefficient
that is only slightly smaller than we find for the numbers of
binaries in the core.  For the 1+2 and especially the 2+2 collision 
rates, however, our results are consistent with little to no 
correlation with BS population size.  The relations for the 
1+1, 1+2, and 2+2 collision rates are plotted in
Figure~\ref{fig:Gamma11_vs_Npop}, Figure~\ref{fig:Gamma12_vs_Npop},
and Figure~\ref{fig:Gamma22_vs_Npop}, respectively, along with 
the corresponding lines of best-fit given in 
Table~\ref{table:least-squares}.

\subsection{Joint Models} \label{joint}

Several authors have suggested that multiple mechanisms could 
contribute significantly to BS formation.  In particular, different 
formation mechanisms could 
operate simultaneously \textit{within the same cluster} 
\citep[e.g.][]{ferraro04}, and/or 
different formation mechanisms could dominate \textit{in different 
clusters} \citep[e.g.][]{davies04}.  One prediction that is 
consistent with the second scenario is that the BSs in the 
high-density clusters in our sample were formed from 
collisions, whereas the BSs in the low-density clusters were formed from 
binary star evolution.  The first scenario, on the other hand, predicts 
that some linear combination of the parameters N$_{bin}$, N$_{1+1}$, 
N$_{1+2}$, and N$_{2+2}$ should yield the strongest correlation with 
observed BS numbers.

We perform two additional comparisons in an effort to test the idea that 
multiple formation mechanisms contributed to the formation of the BSs in 
our sample.  First, 
we divide our sample into low- ($\log \rho_0 < 3.3$) and high-density
($\log \rho_0 > 3.3$) sub-samples of roughly equal size, and
independently re-perform our analysis on each.  This comparison 
will tell us whether different formation mechanisms dominate in 
each of these sub-samples independently.  We use the cluster density 
to divide our sample since, if collisions do contribute to BS formation, 
they should occur with the greatest frequency in high density clusters.

For both sub-samples of low- and high-density clusters, our results 
remain consistent with what 
we found for the entire sample to within the uncertainties.  This is 
supported both by our lines of best-fit and our Spearman correlation 
coefficients.  Therefore, this is consistent with the general 
picture that the dominant BS formation mechanism is the same in all
clusters (as opposed to different mechanisms dominating in
different clusters).

Next, we search 
for a linear combination of the different formation channels that 
yields a better correlation with the observed BS numbers than any 
of these parameters individually.  This is done in two ways.  
Specifically, we fit to the observed data relations of the form:
\begin{equation}
\label{eqn:N_mod2}
N_{BS} = aN_{bin,core} + bN_{1+1} +cN_{1+2} + dN_{2+2},
\end{equation}
and
\begin{equation}
\label{eqn:N_mod3}
N_{BS} = ef_{bin,core}^fM_{core}^g + hf_{bin,core}^iN_{1+1}^j,
\end{equation}
where $e$, $f$, $g$, $h$, $i$, and $j$ are all treated as free 
parameters, and we omit the factor $(1 - f_{bin,core})^{-2}$ when 
calculating $N_{1+1}$ from Equation~\ref{eqn:N_pop_coll2}.  
These comparisons will help to tell us if multiple formation mechanisms 
contribute significantly to BS formation \textit{in the same 
cluster}, or if there is always a dominant formation channel.  

Equation~\ref{eqn:N_mod2} is unable to provide a more statistically 
significant correlation than we find between BS numbers and the 
core masses.  The best-fitting model offers at best a slight 
improvement over what we find upon comparing BS population size 
to either the number of binaries in the core or the 1+1 collision 
rate alone.  Finally, the only best-fitting parameter in 
Equation~\ref{eqn:N_mod3} we 
find to be consistent with a non-zero value is the power-law
index on core mass, with $g = 0.48^{+0.10}_{-0.06}$.  This
independently confirms that the strongest dependence we find
is between BS numbers and core mass.  Therefore, we do not
find evidence from this comparison that multiple mechanisms
contribute simultaneously to BS formation in \textit{individual}
clusters.  

\begin{table}
\caption{Slopes for all weighted least-squares fits, and the
  corresponding Spearman
  rank correlation coefficients and probabilities that the null
  hypothesis of zero correlation is disproved.}  
\begin{tabular}{|c|c|}
\hline
Parameter              &           log N$_{BS}$           \\
 &	slope; r$_s$; p$_s$ \\
\hline
log N$_{bin,core}$     &  0.48 $\pm$ 0.09; 0.61; 3.80e-4  \\
log N$_{bin,half}$     &  0.53 $\pm$ 0.11; 0.49; 5.91e-3  \\
log $\Gamma_{1+1}$     &  0.16 $\pm$ 0.04; 0.60; 5.14e-4  \\
log $\Gamma_{1+2}$     &  0.13 $\pm$ 0.06; 0.36; 4.86e-2  \\
log $\Gamma_{2+2}$     &  0.04 $\pm$ 0.06; 0.06; 7.65e-1  \\
log M$_{core}$         &  0.40 $\pm$ 0.05; 0.83; 1.57e-8  \\
\hline
\end{tabular}
\label{table:least-squares}
\end{table}

\section{Discussion} \label{discussion}

\subsection{Why do the empirical binary fractions fail to yield
  improved correlations for blue stragglers?}

Our main result is that core mass remains a better predictor of blue
straggler numbers than any other variable we have considered. This is
surprising. In the context of the binary evolution scenario for blue
stragglers, one might have expected a stronger correlation with
the number of binaries in the core (as estimated by $f_{bin,core}
\times M_{core}$). Similarly, in the context of the collision scenario, it
seems reasonable to suppose that blue stragglers may form primarily
in collisions involving binaries. In this case, stronger correlations
with either 1+2 or 2+2 collision rates might be expected. Yet neither of 
these hypotheses is confirmed by the data.

At first sight, this negative result is all the more surprising
because the binary fractions themselves {\em do} (anti-)correlate
strongly with core mass (and absolute cluster mass). As noted
in \citet{knigge09}, this anti-correlation is exactly what is needed to
explain the sub-linear correlation between BS numbers and core mass in
the binary evolution model. So why does the combination of core mass and
binary fractions not lead to a roughly linear correlation between
binary and BS numbers, as one might naively expect?

It is, of course, possible that the core mass correlation
is simply the most fundamental one. However, if this correlation is not
driven by binaries, its origin and sub-linear nature are quite hard to
understand. 

As it turns out, there is, in fact, a simple way to understand our
results in the context of the binary evolution model. The easiest way
to see this is to recognize that the existence of a correlation
between $M_{core}$ and $f_{bin,core}$ (Figure~\ref{fig:Mcore_fbin}) 
obviously implies that core mass is itself an estimator of
core binary fractions. Thus even if $N_{BS}$ depends solely on
$N_{bin,core}$, replacing $M_{core}$ with $f_{bin,core} \times
M_{core}$ will only lead to an improved correlation if the
observational errors on the empirical binary fractions are smaller
than the intrinsic scatter in the $f_{bin,core}$ versus $M_{core}$
relationship.

In order to illustrate this quantitatively, we have carried out some
simple simulations. Thus we create mock data sets containing $N = 30$
data points and spanning roughly the same dynamical range as the real
data. In each mock data set, we assume that the true number of BSs 
scales perfectly and linearly with the number of binaries. We also
assume that binary fractions correlate sub-linearly with core masses,
$f_{bin,core} \propto M_{core}^{-0.6}$, and that this correlation is
quite tight, with an intrinsic dispersion of $\sigma_{int} = 0.1$
dex. We also assume 
that BS numbers are subject to an observational error of 0.1 dex, and,
for simplicity, that core masses are perfectly known. Finally, we
assume that the binary fractions are subject to observational
uncertainties, $\sigma_{obs}(f_{bin,core})$. We are interested in how
the character of observationally inferred correlations changes when
$\sigma_{obs}(f_{bin,core})$ approaches and exceeds $\sigma_{int}$, so
we run tests over the range $0.1 \sigma_{int} \leq
\sigma_{obs}(f_{bin,core}) \leq 10.0 \sigma_{int}$. For each trial
value of $\sigma_{obs}(f_{bin,core})$, we create 1000 mock data
sets and measure the Spearman-rank correlation coefficients of the
$N_{BS,obs}$ versus $M_{core}$ relation and the $N_{BS, obs}$ versus 
$N_{bin,obs}$ 
relation. We also measure the slope of the $\log{N_{BS,obs}}$ versus 
$\log{N_{bin,obs}}$ correlation in each mock data set, in order to check if
and when this deviates substantially from the true slope of unity.

\begin{figure}
\begin{center}
\includegraphics[width=\columnwidth]{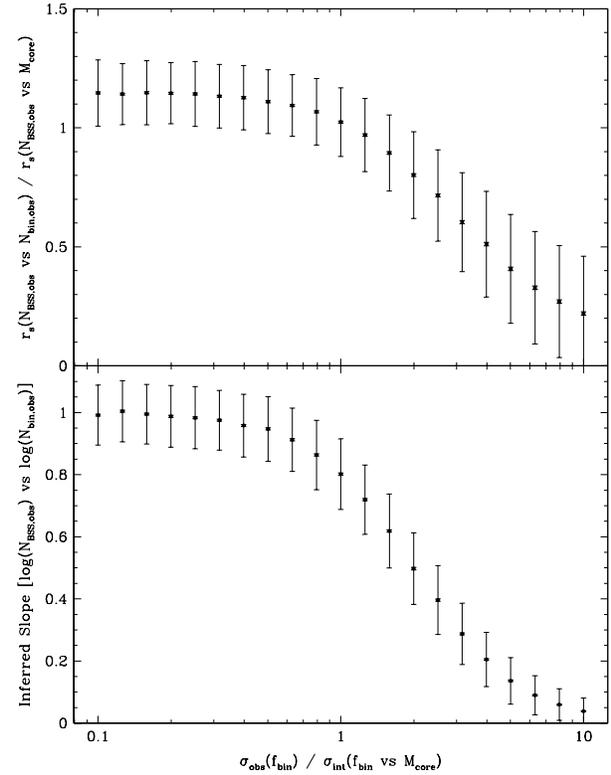}
\end{center}
\caption[Results of our simulations to quantify whether the observational 
errors on the empirical binary fractions are smaller than the intrinsic scatter 
in the $f_{bin,core}$ versus $M_{core}$ relation]{The results 
of our simulations to quantify whether the observational 
errors on the empirical binary fractions are smaller than the intrinsic 
scatter in the $f_{bin,core}$ versus $M_{core}$ relation.  Our procedure 
for this has been described in detail in the text.  Note that the error 
bars shown in this plot correspond to the standard deviation across the 
mock samples, as opposed to the error for the mean.
\label{fig:BLA}}
\end{figure}

The results of the simulations are shown in Figure~\ref{fig:BLA}. They
confirm that the correlation coefficient of the
$N_{BS,obs}$ versus $N_{bin,obs}$ relation exceeds that of the
$N_{BS,obs}$ versus $M_{core}$ relation only if 
$\sigma_{obs}(f_{bin,core}) \lesssim \sigma_{int}$. Once the errors on
$f_{bin,core}$ exceed this, the correlation with core mass becomes
stronger than that with $N_{bin,obs}$, as in the actual data. It is
also interesting that, in the same regime, the measured {\em slope} of
the $\log{N_{BS,obs}}$ versus $\log{N_{bin,obs}}$ relation becomes
significantly shallower than the true slope of unity. Again, this
matches what we see in our analysis of the actual data.

We therefore suggest that the simplest way to understand our results
is to assume that the observational uncertainties on the empirical
binary fractions exceed the intrinsic dispersion in the
$f_{bin,core}$ versus $M_{core}$ relationship. If this is correct, the
observational data can be understood in the context of the simple
binary evolution model. It is worth stressing here that we are not
suggesting that the observed binary fractions are ``wrong'' -- merely
that the uncertainties affecting them are larger than
$\sigma_{int}$. This, in turn, is astrophysically important. The observed
anti-correlation between $f_{bin,core}$ and core mass
(Figure~\ref{fig:Mcore_fbin}) or total cluster mass (\citet{milone11})
is already surprising and should be a key benchmark for dynamical
models of GCs. If our suggestion here is correct, this correlation is
even tighter than the present data suggest.

Having suggested our preferred explanation for our initially
surprising results, we will devote the rest of this section to explore
the viability of other possibilities. In particular, we will carefully
consider the potential impact of key assumptions in our analysis.

\subsection{A check on key assumptions in the analysis}

\subsubsection{Constant average stellar mass}

We have assumed throughout our analysis that the average stellar mass, 
$\bar{m}$, is constant across all clusters. In reality, the average
main-sequence mass can range from $\sim 0.3$ M$_{\odot}$ to 
$\sim 0.7$ M$_{\odot}$, and there seems to be a connection between mass 
function slope and the total GC mass, with lower-mass GCs being more 
depleted of preferentially low-mass stars \citep{leigh12}.  This suggests 
that the average stellar mass should increase with decreasing cluster 
mass.  However, replacing the assumption of a constant average mass in 
Equation~\ref{eqn:N_pop_bin} with a dependence of the form 
$\bar{m} \propto M_{core}^{\epsilon}$, with $\epsilon < 0$, should 
further \textit{flatten} the dependence 
of BS numbers on the calculated numbers of binaries. This suggests 
that our assumption for the average stellar mass is not the cause of the 
observed non-linear dependence of BS numbers on the numbers of binaries.  

We adopt the same assumption of a constant $\bar{m}$ when calculating 
the 1+1, 1+2, and 2+2 collision rates.  The power-law indices 
we find with BS numbers for all three of these parameters are very small 
($\approx 0.1$), and the range in average stellar masses for our sample 
is at most $\approx 0.4$ M$_{\odot}$.  Therefore, there is no realistic 
assumption for the average stellar mass that we could adopt to recover 
a linear relation between BS population size and any of the collision 
rates.  

\subsubsection{Constant average semi-major axis for binaries}

Similarly, we assumed a constant value of $\bar{a} = 2$ AU for 
the average semi-major axes of all binaries undergoing 1+2 and 2+2 
encounters.  However, it is possible that $\bar{a}$ depends 
systematically on the cluster mass, since the semi-major axis 
corresponding to the hard-soft boundary depends on the cluster mass 
(via the velocity dispersion).  We replaced $\bar{a} = 2$ AU in our 
estimates for the 1+2 and 2+2 collision rates with the semi-major 
axis corresponding to the hard-soft boundary in each cluster, and 
re-performed our analysis for these two parameters.  This 
did not improve the agrement between the observed BS numbers and 
either the 1+2 or 2+2 collision rates.  

\subsubsection{Other cluster-to-cluster variations}

There are a number of ways that cluster-to-cluster variations 
in the distributions of binary orbital parameters could have affected 
our results.  For instance, there are several reasons why high-mass
MS-MS binaries with mass ratios $q \sim 1$ are the most likely to produce
BSs.  It is MS stars with masses just below that of the turn-off that
are next in line to ascend the giant branch. Provided they are in 
binaries, they are therefore the best candidates to over-fill their
Roche lobes within the next few hundred Myrs. Two MS stars with
masses close to the turn-off should also produce the brightest and
bluest BSs upon merging \citep[e.g.][]{sills01}, either via binary
coalescence or collisions. These are the most likely to stand out as
BSs in the cluster colour-magnitude diagram.  

The distribution of binary orbital separations should 
affect not only the frequency of mass-transfer events, but also the outcomes
of dynamical interactions involving binaries.  It is the closest binaries 
that are the most likely to tidally interact and undergo mass-transfer
\citep[e.g.][]{mathieu09}, and the probability of a
collision occurring during 1+2 and 2+2 interactions increases with
decreasing binary semi-major axis \citep{fregeau04}. Similar arguments
can also be extended to orbital eccentricity distributions that are
richer in high-eccentricty orbits.  All of this suggests that the
dependences of the various binary parameter 
distributions on total cluster mass could, in principle, play a 
role in driving the observed correlations (or lack thereof).

\subsubsection{The neglect of dynamics}

If a given BS currently resides in the cluster core, this does not
necessarily mean that it formed there.  In particular, many BSs
could have either formed outside the core before migrating in due to
dynamical friction, or they
could have formed inside the core from binary progenitors that recently
migrated in \citep[e.g.][]{mapelli06}.  We found in \citet{leigh11c} that
the observed dependence of BS numbers on core mass can only be reproduced if
at least some BSs did indeed recently migrate in from outside the core.
Specifically, we found that this contributed to lowering the power-law 
index on $M_{core}$.  This is because the time-scale for two-body
relaxation increases with increasing cluster mass, so that fewer BSs
formed outside the cores of more massive clusters have had 
sufficient time to migrate in via dynamical friction.  If correct, this 
predicts that 
the \textit{global} numbers of BSs should correlate more linearly
with the \textit{total} cluster mass than we have found for the relation
between the numbers of BSs in the \textit{core} and the \textit{core} 
masses.  This can be tested directly using a large sample of cluster 
CMDs derived using a field of view that extends out to the tidal radius 
in all clusters.  The on-going work of, for example, 
\citet{fekadu07}, \citet{dalessandro09}, \citet{carraro11}, 
\citet{beccari11}, and \citet{sanna12} should prove very useful in this 
regard.  This is because they have slowly been compiling a large 
sample of CMDs with nearly complete spatial coverage, and it will 
be possible to compile from this a homogeneous sample of cluster CMDs 
for which the field of view consistently includes the entire cluster.

On the other hand, most normal binaries currently populating
the core are likely to have spent a significant fraction of their
lives there \citep[e.g.][]{heggie03}.  This means that they have had
plenty of time to have been affected by the cluster dynamics.
Therefore, any BSs currently in the core that recently migrated in from
the cluster outskirts were formed from a more primordial component of the
total binary population, whereas the present-day core binary fractions
reflect a more dynamically-processed component.  The effects of this
could include a 
weakening of the correlation between the observed number of BSs in the
core and the \textit{present-day} number of binaries in the core.  
However, it would \textit{not} affect an underlying correlation with 
the cluster or core mass.  This would be consistent with the data, 
so this effect may contribute to our results.

\subsubsection{Selection effecs}

One final effect worth considering is the reliance of our BS selection
criteria on location in the colour-magnitude diagram.  In particular, we
find that the number of BSs scales sub-linearly with the core mass.  
Therefore, in order for some issue with our CMD-based selection criteria 
to explain our results, we require that proportionately fewer BSs appear 
considerably brighter and bluer than the main-sequence turn-off in
massive cluster cores when compared to low-mass cores.  This could
arise if, for example, more merger and mass-transfer products appear
hidden along the MS in preferentially massive clusters, instead of
appearing distinctly brighter and
bluer than the MSTO.  If correct, this predicts that the average BS
luminosity should decrease with increasing core mass.  It follows that
the average BS mass should also decrease with increasing core mass, since
previous studies have shown that the luminosities of BSs are correlated
with their masses \citep[e.g.][]{sills01}.  This offers a useful test 
of the idea that the number of ``blue stragglers'' hidden along the 
MS in the CMD depends systematically on the total cluster mass.

\section{Summary}

We have carried out a statistical analysis to study the origins 
of blue stragglers in a large sample of Galactic globular clusters, 
based on data obtained as part of the ACS Survey for Globular
Clusters. The main novel ingredient in our analysis are empirically
estimated core binary fractions, which allow us to estimate the number
of core binaries, as well as the 1+2 and 2+2 collision
rates. Contrary to our expectations, we have found that none of these
observationally estimated quantities yield correlations with BS
numbers that improve upon the previously known sub-linear correlation
between BS numbers and core mass. This is despite the fact that the binary
fractions themselves anti-correlate strongly with core mass, just as
expected in a simple binary evolution model, where the number of BSs would
scale linearly with the number of binaries.

We have explored several possible explanations for our results. The
simplest, and most appealing, is that observational uncertainties
affecting the core binary fractions exceed the intrinsic scatter
of the relationship between binary fractions and core mass. This could
reconcile the data with the binary evolution model. In the context of
the binary evolution model, this would explain why the product of
binary fraction and core mass is a poorer predictor of BS numbers than
core mass alone, and also why the relationship between the observed 
numbers of binaries and BS numbers is sub-linear. 

If this explanation is correct, it would imply that core binary
fractions are tightly coupled to the core or total
cluster mass. This would be of considerable significance for the dynamical 
evolution of globular clusters, and provides an important benchmark 
for simulations attempting to understand their present-day properties.

\section*{Acknowledgments}

We would like to thank Craig Heinke for useful comments, Frank
Verbunt, Tom Maccarone, and Dave Zurek for helpful suggestions for 
improvement, as well as 
an anonymous referee for many suggestions that resulted in
significant improvements to our manuscript.  This
research has been supported by ESA, NSERC and the Ontario Graduate
Scholarship program.

\bsp

\label{lastpage}

\end{document}